\documentclass[runningheads]{llncs}
\usepackage[T1]{fontenc}
\usepackage{graphicx}
\usepackage{wrapfig}
\usepackage{amsmath,amssymb}
\usepackage{algorithm}
\usepackage{algorithmic}
\usepackage{comment}
\usepackage{amsmath,amsfonts,bm}

\def\eqref#1{equation~\ref{#1}}

\def\1{\bm{1}}

\def\va{{\bm{a}}}

\def\vm{{\bm{m}}}

\def\vs{{\bm{s}}}

\def\vx{{\bm{x}}}

\DeclareMathAlphabet{\mathsfit}{\encodingdefault}{\sfdefault}{m}{sl}
\SetMathAlphabet{\mathsfit}{bold}{\encodingdefault}{\sfdefault}{bx}{n}

\newcommand{\E}{\mathbb{E}}

\newcommand{\KL}{D_{\mathrm{KL}}}

\begin{document}
%
\title{Reward-Independent Messaging for Decentralized Multi-Agent Reinforcement Learning}

%
%
\author{Naoto Yoshida\inst{1}\orcidID{0000-0002-9813-0668} \and \\
Tadahiro Taniguchi\inst{1,2}\orcidID{0000-0002-5682-2076} }
\authorrunning{N. Yoshida and T. Taniguchi}
%
\institute{Kyoto University, Kyoto, Japan \and
Research Organization of Science and Technology, Ritsumeikan University \\
\email{\{yoshida.naoto.8x,taniguchi.tadahiro.7j\}@kyoto-u.ac.jp}
}
\maketitle              
\begin{abstract}
In multi-agent reinforcement learning (MARL), effective communication improves agent performance, particularly under partial observability. We propose MARL-CPC, a framework that enables communication among fully decentralized, independent agents without parameter sharing. MARL-CPC incorporates a message learning model based on collective predictive coding (CPC) from emergent communication research. Unlike conventional methods that treat messages as part of the action space and assume cooperation, MARL-CPC links messages to state inference, supporting communication in non-cooperative, reward-independent settings. We introduce two algorithms—Bandit-CPC and IPPO-CPC—and evaluate them in non-cooperative MARL tasks. Benchmarks show that both outperform standard message-as-action approaches, establishing effective communication even when messages offer no direct benefit to the sender. These results highlight MARL-CPC’s potential for enabling coordination in complex, decentralized environments.

\keywords{Multi-agent Reinforcement Learning \and Emergent Communication \and Predictive Coding \and Collective Predictive Coding}
\end{abstract}

\section{Introduction}

In multi-agent reinforcement learning (MARL), inter-agent communication for sharing private observations can yield group-level benefits \cite{zhu2024survey,gronauer2022multi,albrecht2024multi}. Under partial observability, agents can improve decision-making by integrating localized perceptual inputs. Such communicative behaviors are common in nature; for example, vervet monkeys use alarm calls to warn conspecifics of threats, enabling collective risk avoidance beyond the perceptual range of any individual \cite{seyfarth1980monkey}. Similarly, in human language, the exchange of individual experiences facilitates knowledge integration, supporting collective adaptation and utility \cite{nowak1999evolution,tomasello2010origins}.

Natural agents that engage in communication function as autonomous units with decentralized learning mechanisms. Reinforcement signals (rewards) are assigned individually and may not be aligned, making the environment inherently non-cooperative. As a result, effective communication must emerge under decentralized and potentially conflicting incentive structures. Despite its importance for both practical MARL applications \cite{pina2024fully,lin2021learning,wong2023deep} and foundational research on emergent communication \cite{lazaridou2020emergent}, the challenge of enabling functional communication among independently learning, decentralized agents remains underexplored.

The main contribution of this study is the proposal of MARL-CPC, a novel deep MARL framework that enables independent agents to establish and utilize communication through decentralized learning. Building on the concept of Collective Predictive Coding (CPC)—recently introduced in emergent communication research~\cite{taniguchi2024collective}—we develop two algorithms: Bandit-CPC and IPPO-CPC. These methods facilitate communication among independently optimizing agents, regardless of cooperative conditions. Empirical evaluations demonstrate that MARL-CPC significantly improves group-level performance by enabling effective information sharing, even in non-cooperative scenarios.

\section{Related Work}
\subsection{MARL with Communication}

Communication has been shown in numerous studies to enhance performance in MARL \cite{gronauer2022multi,zhu2024survey,wong2023deep}. In many cases, such optimization relies on two key assumptions.

First, from an engineering-oriented perspective, centralized training with decentralized execution (CTDE) has been widely adopted, linking agents through a centralized optimization framework \cite{sukhbaatar2016learning}. This includes the use of global value functions \cite{sunehag2018value,lowe2017multi,yu2022surprising} and architectures like RIAL, DIAL \cite{foerster2016learning}, and CommNet \cite{sukhbaatar2016learning}, which allow gradient propagation across agents to improve policy optimization \cite{zhu2024survey}. Parameter sharing is also commonly employed to facilitate learning from individual experiences \cite{zhu2024survey}. However, these centralized approaches often diverge from the fully decentralized learning observed in natural agents. They typically assume unrealistic access to shared information and coordination mechanisms, limiting their relevance to models of emergent communication. Additionally, reliance on agent homogeneity has been criticized for its inability to support role differentiation based on context \cite{wong2023deep}.

Second, most prior work on communication in MARL assumes cooperative environments \cite{lazaridou2018emergence}. Insights from simulation studies on language evolution suggest that in systems of independently learning, decentralized agents with individual objectives, communication is often hindered by free-riding and deceptive signaling, impeding the stability of shared communication protocols \cite{Mirolli2010}. Similar challenges arise in the “cheap-talk” framework from economics \cite{farrell1996cheap}, where communication is costless and non-binding. Consequently, MARL studies using this framework typically adopt cooperative settings such as signaling games \cite{lewis1969convention,skyrms2010signals,ueda2024lewis} or referential games \cite{lazaridou2017multi}.

Given these challenges, the emergence of communication among decentralized, independently learning agents in non-cooperative environments remains difficult. Under such conditions, it is reasonable to assume that introducing a communication-inducing module is necessary for facilitating communication. While several studies have explored decentralized acquisition of such modules \cite{lin2021learning,ebara2023multi,nakamura2023control,pina2024fully}, these efforts primarily focus on cooperative settings. In this study, we investigate the possibility and implications of communication emerging under non-cooperative conditions.

\subsection{Collective Predictive Coding}

{\it Collective Predictive Coding} (CPC) is an emergent communication model for independent, decentralized agents, originally proposed in the field of emergent communication research \cite{taniguchi2023emergent,hoang2024emergent,taniguchi2024collective}. It extends predictive coding theory from computational neuroscience \cite{rao1999predictive,friston2006free} to multi-agent systems. In developmental psychology, human communication that benefits others is known to exhibit altruistic properties in natural environments \cite{tomasello2009we}.
CPC assumes that human linguistic behavior and its acquisition are driven by innately altruistic mechanisms, drawing inspiration from the cognitive and motor development processes involved in human language learning \cite{taniguchi2024collective}. 

Rather than modeling communication emergence through RL, CPC formulates it as inference within a single, large generative model representing a group of agents. This model is decomposed across individuals, yielding an objective function that each agent can optimize via communication \cite{hoang2024emergent,taniguchi2024collective}.
This formulation enables decentralized Bayesian inference in a distributed manner \cite{hagiwara2019symbol,taniguchi2023emergent,nomura2025decentralized}. In this context, messages are interpreted as auxiliary variables supporting distributed optimization. Viewing the multi-agent system as a single generative model also allows communication learning to be understood as a process of knowledge integration based on individually acquired observations.

Previous applications of CPC to MARL have been explored by Ebara et al. \cite{ebara2023multi} and Nakamura et al. \cite{nakamura2023control}, but their approaches rely on posterior sampling via Markov Chain Monte Carlo (MCMC), limiting compatibility with neural network-based function approximation. In contrast, the present study introduces a CPC formulation grounded in variational inference, enabling implementation with deep neural networks and thus offering greater scalability and practical applicability.


\section{Multi-Agent Reinforcement Learning with Collective Predictive Coding}
\subsection{Preliminaries}
The problem addressed in this study is formulated as a Partially Observable Markov Game (POMG) \cite{hansen2004dynamic}. Formally, a POMG is defined by the tuple \\$\left<\mathcal{I}, \mathcal{S}, \mathcal{A}, \mathcal{X}, \mu_0, P, R \right>$, where $\mathcal{I}$ denotes the set of $N$ agents; $\mathcal{S}$, the set of environmental states; and $\mathcal{A} = \mathcal{A}_1 \times \mathcal{A}_2 \times \dots \times \mathcal{A}_N$, the joint action space composed of each agent $i$’s action set $\mathcal{A}_i$. Similarly, $\mathcal{X} = \mathcal{X}_1 \times \mathcal{X}_2 \times \dots \times \mathcal{X}_N$ denotes the joint observation space. The initial state distribution is denoted by $\mu_0 \in \triangle(\mathcal{S})$, where $\triangle(\cdot)$ represents the set of probability distributions over the given set. The transition-observation function is given by $P(\vs', \vx \mid \vs, \va)$, where $s' \in \mathcal{S}$ is the next state, $\vx = (x_1, x_2, \dots, x_N) \in \mathcal{X}$ is the joint observation, and $\va = (a_1, a_2, \dots, a_N) \in \mathcal{A}$ is the joint action. The reward function for agent $i$ is defined as $R_i: \mathcal{S} \times \mathcal{A} \rightarrow \mathbb{R}$.

In a POMG, each agent independently aims to maximize its expected return, $\mathbb{E}{\pi}\left[\sum_{t=0}^\infty \gamma^t r_{i,t}\right]$, based on its own experiences. Here, $r_{i,t}$ is the reward received by agent $i$ at time step $t$, $\pi_i$ is the policy of agent $i$, and $\mathbb{E}_{\pi}[\cdot]$ denotes the expectation over trajectories induced by the joint policy $\pi = (\pi_1, \pi_2, \dots, \pi_N)$ under the environment dynamics.

To formalize effective communication in this study, we consider a setting within the POMG framework where the state space is factorized as $\mathcal{S} = \mathcal{S}_1 \times \mathcal{S}_2 \times \dots \times \mathcal{S}_N$. We assume $\mathcal{X}_i \triangleq \mathcal{S}_i$ for all $i \in \mathcal{I}$, meaning that each agent observes only its own component of the state. Thus, $x_i$ represents private information for agent $i$, and access to other agents’ observations $x_j$ (for $j \neq i$) effectively reveals the full environmental state $s$, which may enhance the agent's expected future return. We specifically focus on non-cooperative reward settings, where rewards differ across agents: $R_i(\vs, \va) \neq R_j(\vs, \va)$ for some $(\vs, \va) \in \mathcal{S} \times \mathcal{A}$ \cite{busoniu2008comprehensive}, while assuming that agents are not in direct competition. We investigate the emergence and role of communication under these conditions.

\subsection{Variational CPC by Joint Auto-encoder}

\begin{wrapfigure}[15]{r}{0.4\textwidth}
    \centering
    \vspace{-8mm}
    \includegraphics[width=0.4\textwidth]{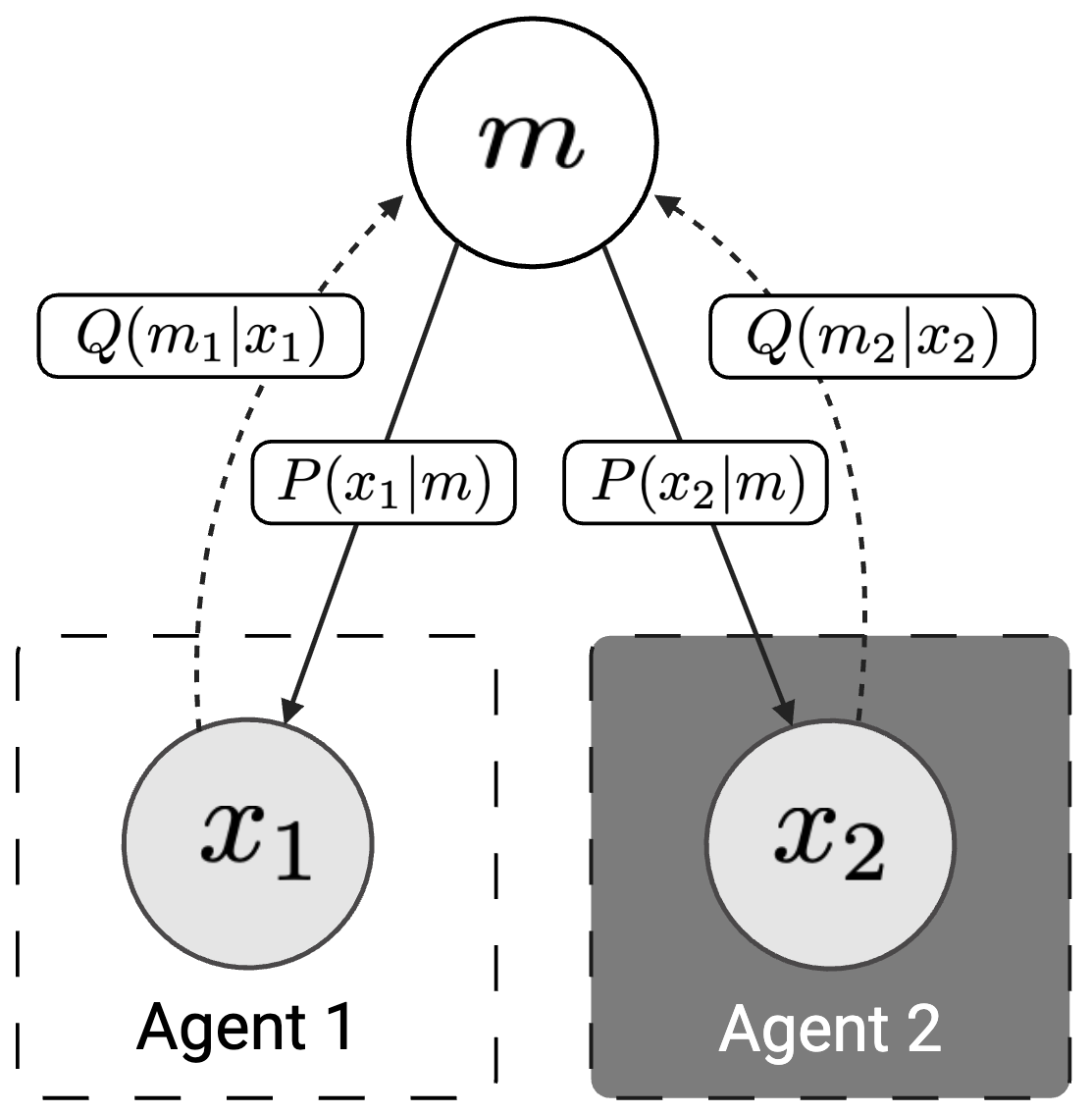}
    \vspace{-7mm}
    \caption{Graphical model of the CPC module (2 agents).}
    \label{fig:gm}
\end{wrapfigure}

This study formulates CPC using variational inference within a deep generative modeling framework. In this context, a joint generative model is constructed by aggregating the observations \(x_i\) of individual agents (\(i = 1, 2, \dots, N\)). This joint model is then decomposed to derive an objective function for the communication modules of individual agents.

An overview of the generative model employed in this study is provided in Figure~\ref{fig:gm}. The model defines the joint probability distribution over all agents’ observations \(x_i\) and messages \(m_i \in \{1, \dots, K\}\) as follows:
\begin{eqnarray}
    P_\theta(\vx, \vm) \triangleq P(\vm) \prod_{i=1}^N P_{\theta_i}(x_i|\vm),
\end{eqnarray}
where $\vx = (x_1, x_2, \dots, x_N)$ and $\vm = (m_1, m_2, \dots, m_N)$ is the joint message.
We consider performing variational inference by deriving an evidence lower bound (ELBO) of the above joint probability, as in variational autoencoders (VAE)~\cite{kingma2013auto}. Let \( Q(\vm) \) denote the variational distribution over the joint message \( \vm \). Then, the likelihood of the observations can be lower-bounded as follows:
\begin{eqnarray}
    \log P_\theta(\vx) &\geq& \sum_\vm Q(\vm) \log \frac{P_\theta(\vx|\vm)P(\vm)}{Q(\vm)}.
\end{eqnarray}
Here, we introduce the variational distribution and the prior distribution over \( \vm \) as follows, and substitute them into the ELBO above:
\begin{eqnarray}
    &Q(\vm) \triangleq Q_{\phi}(\vm|\vx) = \prod_i Q_{\phi_i}(m_i|x_i)&\\
    &P(\vm) \triangleq \prod_{i=1}^N P(m_i)&
\end{eqnarray}
As a result, the ELBO can be decomposed into agent-wise terms as follows:
\begin{eqnarray}
    \sum_\vm Q(\vm) \log \frac{P_{\theta}(\vx|\vm)P(\vm)}{Q(\vm)} 
    &=& \sum_{i=1}^N \left[\sum_\vm Q_{\phi}(\vm|\vx)  \log  \frac{P_{\theta_i}(x_i|\vm)P(m_i)}{Q_{\phi_i}(m_i|x_i)}\right].
\end{eqnarray}
This suggests that the ELBO in CPC can be decomposed with respect to each individual agent \( i \). By expressing each agent's term as 
\begin{eqnarray}
    J_\text{CPC}(\theta_i, \phi_i) &\triangleq& \E_{Q_{\phi}(\vm|\vx)} \left[ \log P_{\theta_i}(x_i|\vm)\right] - \KL\left(Q_{\phi_i}(m_i|x_i)\|P(m_i)\right),
        \label{eq:loss_cpc}
\end{eqnarray}
where $\KL$ represents the Kullback–Leibler divergence. Then we obtain
\begin{eqnarray}
    \log P_\theta(\vx) &\geq&  \left(\sum_{i=1}^N J_\text{CPC}(\theta_i, \phi_i)\right).
\end{eqnarray}
Therefore, by maximizing \( J_\text{CPC}(\phi_i, \theta_i) \) for each agent, the variational distribution \( Q_\theta(\vm \mid \vx) \) approximates the posterior \( P_\theta(\vm \mid \vx) \), and the message variable \( \vm \), which integrates the observations \( \vx \) from both agents, corresponds to a state estimation of the entire environment $s\in \mathcal{S}$. Furthermore, \( Q_{\phi}(\vm \mid \vx) \) in \( J_\text{CPC}(\phi_i, \theta_i) \) can be interpreted as sampling based on the utterances of all agents. In this study, we use the following one-sample approximation:
\begin{eqnarray}
    &J_\text{CPC}(\theta_i, \phi_i) \approx \log P_{\theta_i}(x_i|\vm) - \KL\left(Q_{\phi_i}(m_i|x_i)\|P(m_i)\right)&\\
    &\vm \sim Q_{\phi}(\vm|\vx).&
\end{eqnarray}

\begin{figure}[t]
\centering
\includegraphics[width=\textwidth]{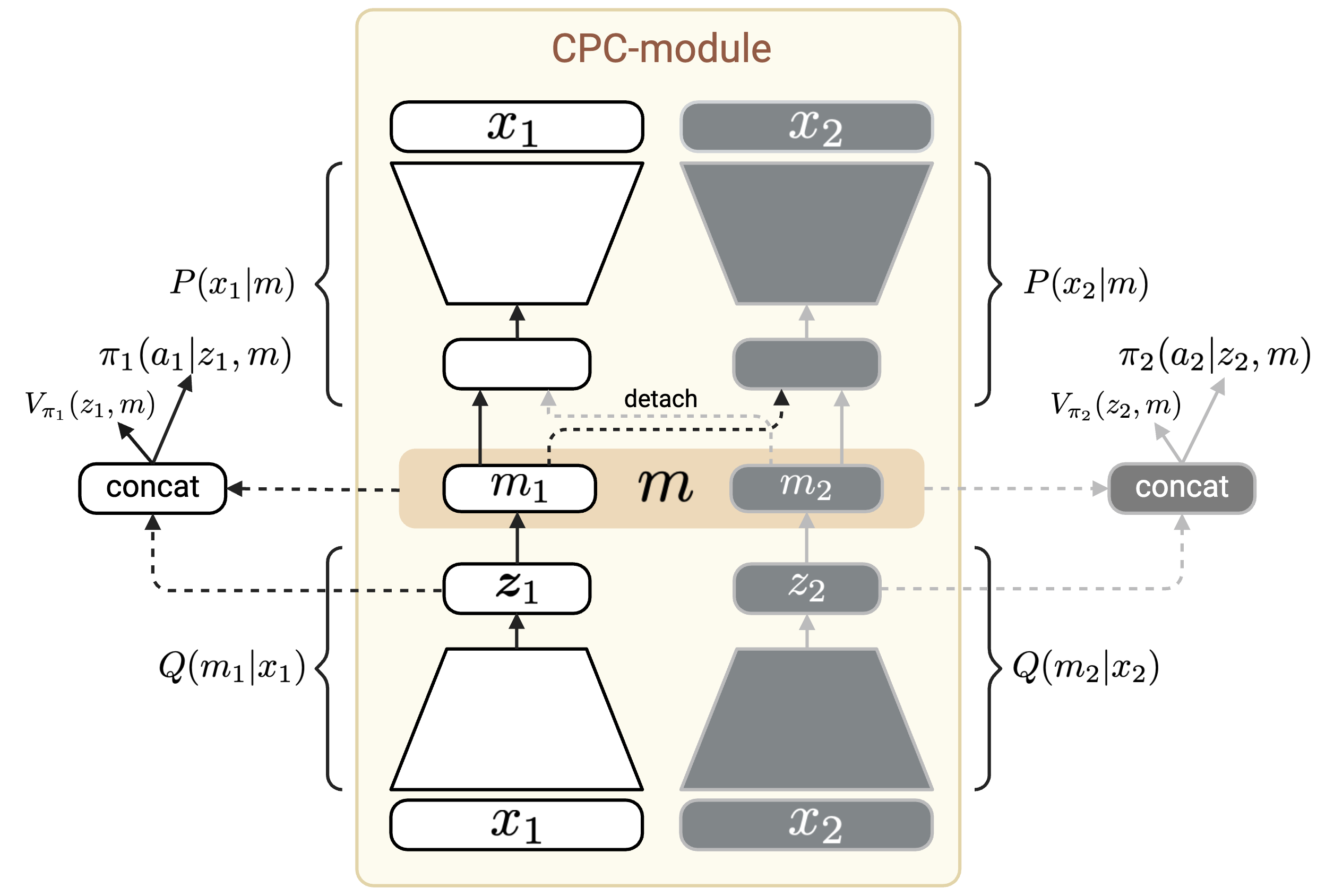}
\caption{Overview of the MARL-CPC architecture.
The figure is a model with two agents. The components of each agent are represented by filled regions—white and gray, respectively. The central panel corresponds to the CPC module, which forms a pseudo-joint agent and enables message generation and exchange. Based on the messages $\vm$ and the hidden states $z$ acquired through the CPC module, the agent performs action selection and value estimation. The dashed arrows in the figure indicate paths through which gradients do not propagate during learning.
} 
\label{fig:architecture}
\end{figure}

The optimization of \( J_\text{CPC}(\phi_i, \theta_i) \) is performed independently by each individual agent. Practically, each agent’s discrete message \( m_i \) is represented as a one-hot feature vector, and \( \vm = (m_1, \dots, m_N) \) denotes the concatenation of messages from \( N \) agents, forming a vector of dimension \( N \times K \). For optimization, we use the following straight-through gradient estimator \cite{bengio2013estimating} for each agent’s own message:
\begin{eqnarray}
    {\tilde m}_i(x_i) = m_i + \log Q_{\phi_i}(m_i|x_i) - \mathbf{sg}\left[\log Q_{\phi_i}(m_i|x_i) \right]
\end{eqnarray}
Here, \( \mathbf{sg}[\cdot] \) denotes the stop-gradient operator. Accordingly, in the training of agent \( i \), gradients are computed using the input \( \vm(x_i) = (m_1, \dots, \tilde{m}_i(x_i), \dots, m_K) \), where only the agent’s own message is treated as a function of its input and the rest are detached. In addition, the KL divergence is approximated using a sampling-based technique inspired by methods from deep RL \cite{schulman2020apporx}, as follows:
\begin{eqnarray}
    &\KL\left(Q_{\phi_i}(m_i|x_i)\|P(m_i)\right) \approx \left(\kappa - 1\right) - \log \kappa,&\\
    &\kappa = \frac{Q_{\phi_i}(m_i|x_i)}{P(m_i)}.&
\end{eqnarray}

The CPC-based communication module resembles the autoencoder-based method proposed by Lin et al.~\cite{lin2021learning}. However, our approach differs in that the decoder reconstructs from the entire message vector \(\vm\), and each agent’s objective function includes an additional KL divergence term. When the KL term is weighted as in a \(\beta\)-VAE~\cite{higgins2017beta} with \(\beta = 0\), and the decoder is constrained such that \(P_{\theta_i}(x_i \mid \vm) \triangleq P_{\theta_i}(x_i \mid m_i)\), our formulation reduces to that of Lin et al., indicating that our method generalizes their approach. Furthermore, whereas the effectiveness of message learning based on autoencoders was not theoretically justified in prior work, our CPC-based derivation interprets the joint message as supporting state estimation, thus providing a principled account of inter-agent information sharing.

\subsection{MARL-CPC}


This study proposes a MARL framework that facilitates communication learning in MARL via a CPC module, termed \textbf{MARL-CPC}. Based on this framework, we introduce two algorithms. The first, \textbf{Bandit-CPC}, is designed for multi-agent contextual bandit problems. The second, \textbf{IPPO-CPC}, extends the approach to more complex scenarios involving state transitions.

During execution, each agent samples a message \( m_i \) from the variational distribution \( Q_{\phi_i}(m_i \mid x_i) \). The concatenated global message \( \vm \), together with each agent’s hidden representation \( z_i \) obtained by embedding its observation \( x_i \), is then provided as input to the subsequent RL modules (Figure~\ref{fig:architecture}).

The objective function of both algorithms is expressed as the sum of the RL term $J_\text{RL}$ and the CPC term $J_\text{CPC}$. The overall objective function to maximize for agent $i$ of MARL-CPC is expressed as follows:
\begin{eqnarray}
    J(\eta_i, \theta_i, \phi_i) = \E_t\left[J_\text{RL}(\eta_i) + J_\text{CPC}(\theta_i, \phi_i)\right],
\end{eqnarray}
Here, \( \eta_i \) denotes the parameters of agent \( i \)'s policy \( \pi_i \). The expectation operator \( \mathbb{E}_t \) represents an empirical average computed over a finite batch of samples. The pseudocode is provided in Algorithm~\ref{alg:marl-cpc}. During RL optimization, gradients are not propagated through the CPC module. Instead, gradient computations for the CPC and RL components are performed independently.

\begin{algorithm}[t]
\caption{MARL-CPC pseudocode}
\begin{algorithmic}[1] 
\STATE Initialize parameters $(\eta_i, \theta_i, \phi_i)$ for each agent.
\FOR{$\text{iteration}=0,1,2,\dots$}
    \STATE{Collect sample sets $\{\mathcal{D}_i\}_{i=1,\dots,N}$ in the environment using $\{\pi_i, Q_{\phi_i}\}_{i=1,\dots,N}$}.
    \FOR{$i=0,1,2,\dots,N$}
        \STATE{Make mini-batches $\{\mathcal{D}_i^k\}_{k=0,1,\dots}$ using data $\mathcal{D}_i$}
        \FOR{all mini-matches}
            \STATE{Calculate the RL loss $J_\text{RL}(\eta_i)$ using data $\mathcal{D}_i^k$ and (\ref{eq:loss_bandit}) or (\ref{eq:loss_ppo}).}
         \STATE{Calculate the CPC loss $J_\text{CPC}(\theta_i, \phi_i)$ using data $\mathcal{D}_i^k$ and (\ref{eq:loss_cpc}).}
        \STATE{Update parameters using some gradient ascent: \begin{eqnarray*}
            (\eta_i, \theta_i, \phi_i)\leftarrow (\eta_i, \theta_i, \phi_i) + \alpha \nabla_{\eta_i, \theta_i, \phi_i}J(\eta_i, \theta_i, \phi_i).
        \end{eqnarray*}}
        \ENDFOR
    \ENDFOR
\ENDFOR
\end{algorithmic}
\label{alg:marl-cpc}
\end{algorithm}

\vspace{-3mm}
\subsubsection{Bandit-CPC}
This algorithm is effective in environments where each agent is provided with individually defined observations and rewards in a contextual bandit setting, and where agents can potentially benefit from sharing information with one another. In this context, letting $r_i$ denote the reward obtained by agent $i$ in a single trial, the objective function to be maximized for improving the policy $\pi_i$ is given as follows.
\begin{eqnarray}
    J_\text{RL}(\eta_i) = r_i \log \pi_i(a_i|z_i, \vm)
    \label{eq:loss_bandit}
\end{eqnarray}
Here, $z_i$ is the internal representation of the encoder $Q(m_i|x_i)$ (Figure~ \ref{fig:architecture}). 

\vspace{-3mm}
\subsubsection{IPPO-CPC}
For environments beyond contextual bandits, we incorporate the CPC module into Proximal Policy Optimization (PPO)~\cite{schulman2017proximal} for representation learning. PPO is an on-policy, actor-critic deep RL method that optimizes a policy \( \pi_i \) and value function \( V_{\pi_i} \), both parameterized by deep neural networks and trained via gradient-based methods~\cite{kingma2014adam}. Independent PPO (IPPO) extends PPO to multi-agent settings, allowing each agent to optimize its policy independently using local updates~\cite{de2020independent}. We denote the combined parameters of agent \( i \)'s policy and value networks as \( \eta_i \). In our implementation, \( \pi_i \) and \( V_{\pi_i} \) are modeled using separate multilayer perceptrons (MLPs).

Let \( \pi_\text{old} \) denote the policy at the time of sampling, and define \( r_t(\eta_i) = \frac{\pi_i(a_t|z_{i,t}, \vm_t)}{\pi_\text{old}(a_t|z_{i,t}, \vm_t)} \). The subscript \( z_{i,t} \) denotes the internal representation of agent \( i \)'s encoder $Q(m_i|x_i)$ at time step \( t \). The PPO objective is expressed as the minimization of the following loss function:
\begin{eqnarray}
    J_\text{RL}(\eta_i) &=& J_\pi(\eta_i) - c_1 J_V(\eta_i) + c_2 H_\pi(\eta_i) \label{eq:loss_ppo}\\
    J_\pi(\eta_i) &=& \min \Bigl( r_t(\eta_i)A_{i, t} , \text{clip}\left(r_t(\eta_i), 1-\epsilon, 1+\epsilon)A_{i, t} \right) \Bigr)\\
    J_V(\eta_i) &=& \left(V_{\pi_i}(z_{i, t}, \vm_t) - V_{i, t}^\text{targ}\right)^2
\end{eqnarray}
Here, the advantage \( A_{i,t} \) is computed using the value estimator at sampling time, based on generalized advantage estimation \cite{Schulmanetal_ICLR2016}. \( H_\pi \) denotes the entropy of the policy. \( c_1=0.5 \), \( c_2=0.01 \), and \( \epsilon=0.2 \) are all positive constants. The function \( \text{clip}(\cdot, 1 - \epsilon, 1 + \epsilon) \) restricts its input to the range \( [1 - \epsilon, 1 + \epsilon] \) \cite{schulman2017proximal}. The target value \( V_t^\text{targ} \) is the empirical return computed from a trajectory of length \( T \), defined as $V_t^\text{targ} = r_t + \gamma r_{t+1} + \cdots + \gamma^{T - t} V_{\pi_i}(z_{i,T}, \vm_T)$.

\vspace{-3mm}
\section{Experiments}
\vspace{-3mm}

\begin{figure}[t]
\centering
\includegraphics[width=\textwidth]{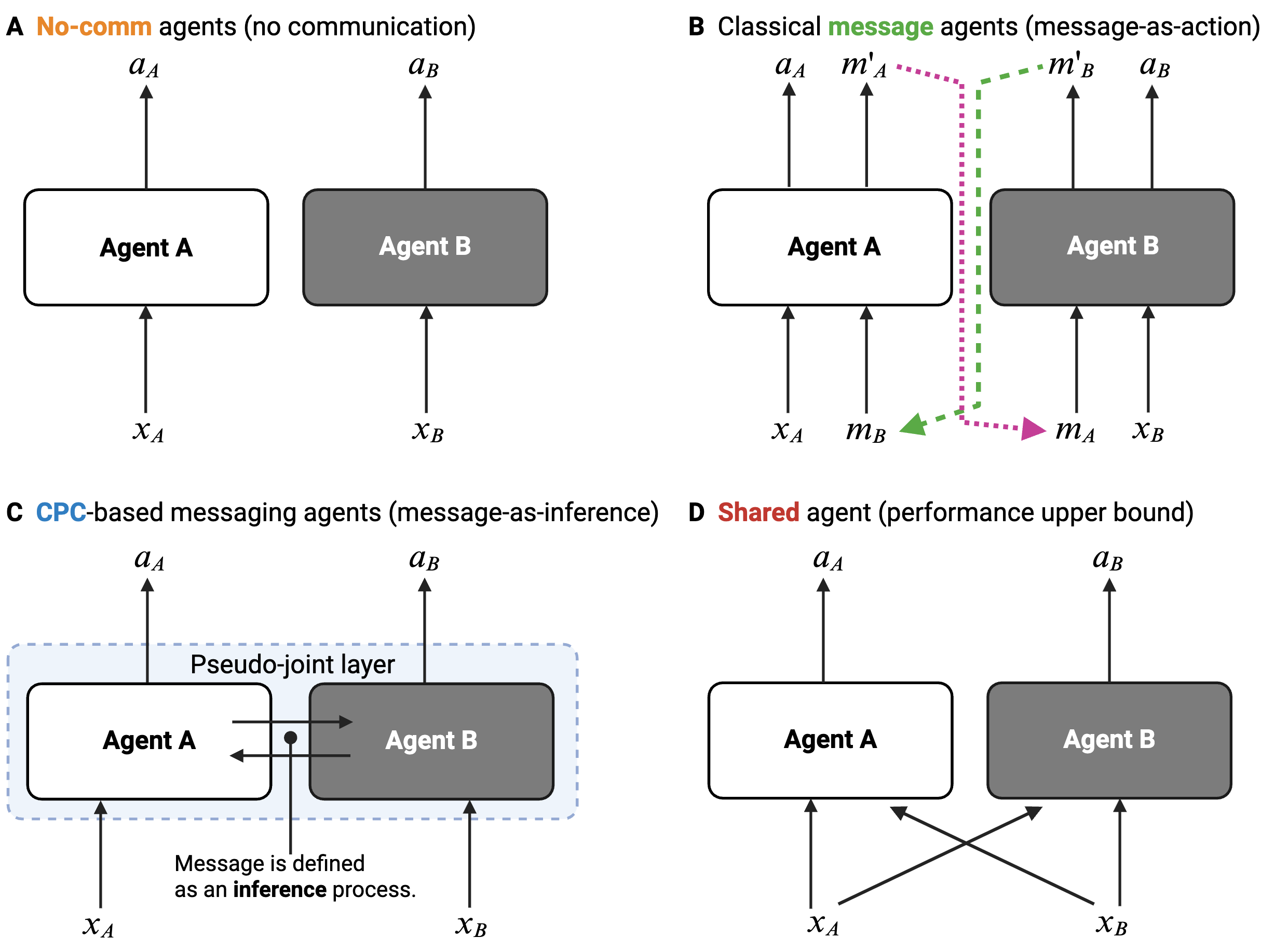}
\caption{Agent architectures compared in this experiments. {\bf A}) Independent agents without communication \cite{de2020independent}. {\bf B}) Message agents, where communication is defined as an extension of action \cite{cangelosi1998emergence,foerster2016learning}. {\bf C}) CPC-based agents in which messages function as auxiliary variables for the state inference process (ours). {\bf D}) Agents whose observations are shared in advance (performance upper bound).} 
\label{fig:message}
\end{figure}
We constructed MARL environments where information sharing via communication influences performance and compared different agent architectures (Figure~\ref{fig:message}). Two environments were evaluated: a multi-agent contextual bandit and the ``observer'' environment. We tested four agent types: independent agents without communication ({\bf no-comm}); agents using messages as actions ({\bf message}), as in classical studies~\cite{cangelosi1998emergence,foerster2016learning}; and agents with CPC-based communication ({\bf cpc}). To estimate the performance upper bound under full information sharing, we also evaluated a {\bf shared} condition, where each agent’s policy and value networks receive both observations \( (x_1, x_2) \) as joint input.

\subsubsection{Implementation details}
For all agent architectures, the policy and value networks consisted of multilayer perceptrons (MLPs) with two hidden layers of 64 units and \texttt{Tanh} activation. In the {\bf message} condition, message inputs and outputs were added to the policy network, while the {\bf cpc} condition incorporated a CPC module. Specifically, \( P_{\theta_i}(x_i \mid \vm) \) and \( Q_{\phi_i}(m_i \mid x_i) \) were implemented as MLPs with a single 64-unit hidden layer and GELU activation~\cite{hendrycks2016gelu}. For message prior $P(m_i)$ we used a flat prior in both experiments. For agents with communication, messages were used as additional inputs to the value network. We used the Adam optimizer with a learning rate of \( 3 \times 10^{-4} \) for Bandit-CPC and \( 2.5 \times 10^{-4} \) for IPPO-CPC. The discount factor for IPPO-CPC was set to \( \gamma = 0.99 \).

\begin{figure}[t]
\centering
\includegraphics[width=0.9\textwidth]{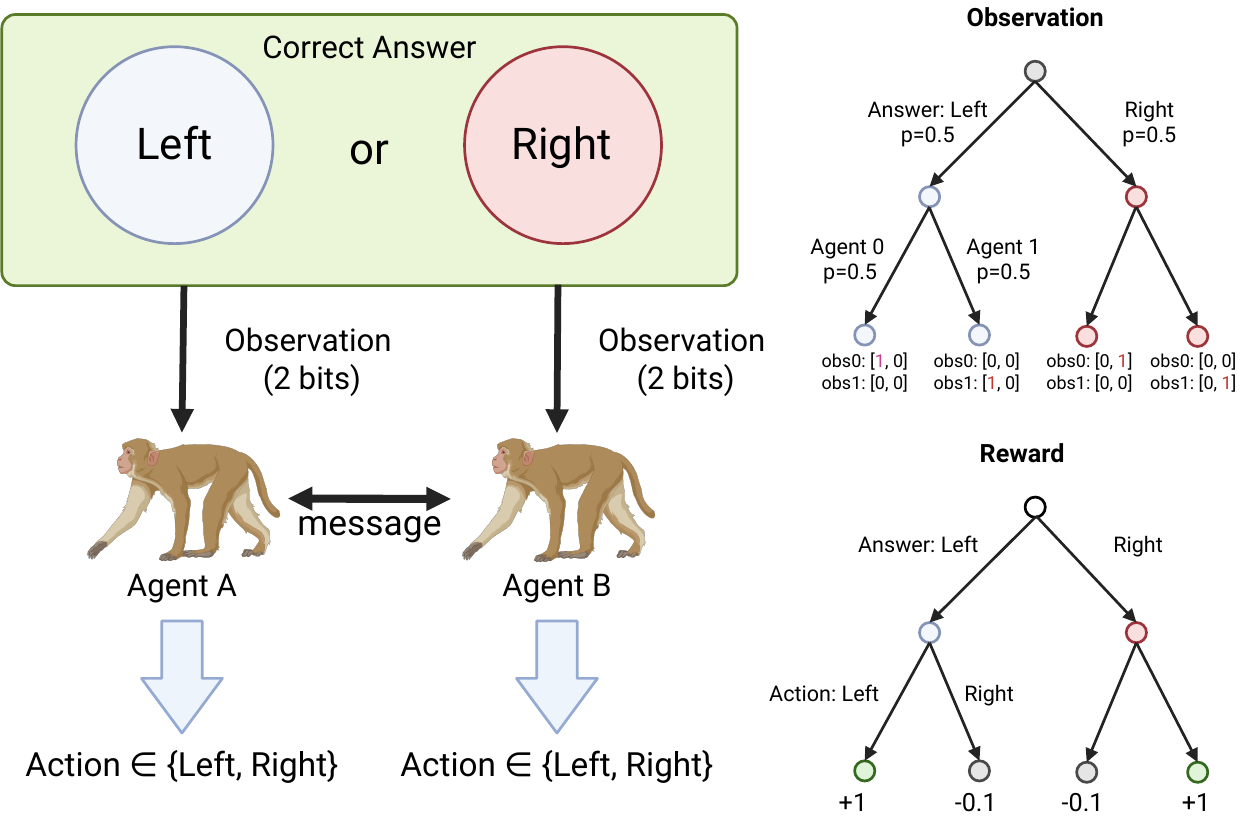}
\caption{Multi-agent conditional bandit environment.} 
\label{fig:envs_bandit}
\end{figure}

\subsection{Contextual Bandit with Information Sharing}

This experiment serves as a proof of concept for communication via CPC in a non-cooperative setting where communication benefits each individual agent (Figure~\ref{fig:r_bandit}). The environment consists of two independently acting agents (Agent-A and Agent-B), each making a single decision per episode and receiving an individual reward based on the environmental state and their chosen action.

Each episode proceeds as follows. The environment has a true state \( s \in \{\text{LEFT}, \text{RIGHT}\} \), sampled uniformly. Only one agent observes the true state, encoded as a binary vector: the informed agent receives \( x_i = [1, 0]^\top \) if \( s = \text{LEFT} \), and \( x_i = [0, 1]^\top \) if \( s = \text{RIGHT} \); the uninformed agent receives \( x_i = [0, 0]^\top \). Each agent selects an action \( a_i \in \{\text{LEFT}, \text{RIGHT}\} \), and receives \( +1 \) if \( a_i = s \), and \( -0.1 \) otherwise.

This setting is non-cooperative: the informed agent can maximize its own reward without relying on the other, and thus has no incentive to communicate. As a result, reward-based learning alone does not promote communication. However, for the uninformed agent, accessing the true state is critical, and shared communication can increase the total group reward. CPC has the potential to establish such communication autonomously.

We evaluated Bandit-CPC under the {\bf cpc} condition and compared it with three baselines: {\bf independent}, {\bf message}, and {\bf shared}. Each condition was run for \( 3.0 \times 10^4 \) episodes. Messages were discrete (5 values), and each agent could send one message per timestep. Performance was measured using group welfare \( W = r_0 + r_1 \)~\cite{albrecht2024multi}, which approaches 2.0 when cooperative communication is achieved.

\vspace{-5mm}
\subsubsection{Results}
\begin{wrapfigure}[10]{r}{0.5\textwidth}
\vspace{-13mm}
\begin{center}
\includegraphics[width=0.5\textwidth]{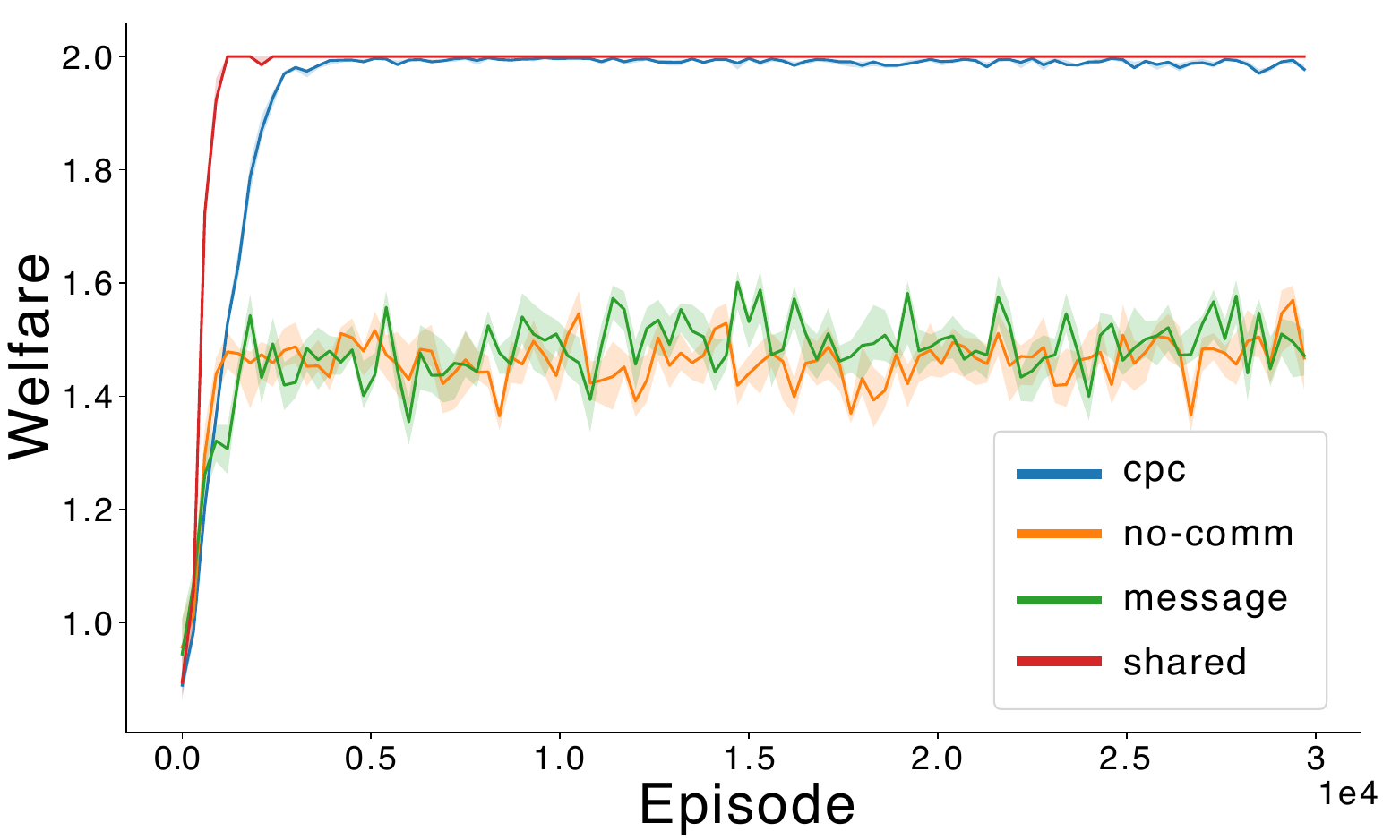}
\end{center}
\vspace{-7mm}
\caption{Results in Bandit environment.} 
\label{fig:r_bandit}
\end{wrapfigure}

Figure~\ref{fig:r_bandit} presents the experimental results. The results are reported as the interquartile mean (IQM) in $N$ runs, accompanied by bootstrapped 95\% confidence intervals (CIs) with 2,000 iterations of resampling \cite{agarwal2021deep}. First, the shared condition (information sharing is assumed in advance) achieves a group welfare close to 2.0, confirming that this is the maximum achievable value under the given learning conditions. 
As shown in the results, agents with the CPC module also attain comparable levels of group welfare, indicating that information sharing is successfully established and utilized by each agent without the need for an explicitly cooperative setting. 

In contrast, message agents fail to establish effective information sharing in this non-cooperative environment, resulting in group welfare that remains at the same suboptimal level as in the independent condition. These results demonstrate that MARL-CPC enables the emergence of beneficial communication even in non-cooperative environments.

\subsection{Observer: Information Sharing without Rewards}

\begin{wrapfigure}[13]{r}{0.5\textwidth}
\vspace{-12mm}
\begin{center}
\includegraphics[width=0.5\textwidth]{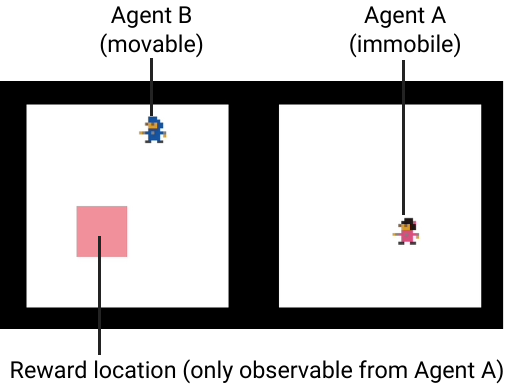}
\end{center}
\vspace{-7mm}
\caption{Overview of observer environment.} 
\label{fig:env_observer}
\end{wrapfigure}

This environment evaluates communication in a non-cooperative setting with asymmetric information access (Figure~\ref{fig:env_observer}). It features two agents: Agent-A, who remains stationary and receives no reward, and Agent-B, who navigates a \( 4 \times 4 \) grid and can earn \( +1 \) by selecting the \texttt{DIG} action on cells with buried rewards. All other actions yield a penalty of \( -0.01 \). Agent-B cannot observe the reward locations, while Agent-A can, but has no incentive to share this information, as it receives zero reward regardless of its actions.

Agent-A's observation is a 16-dimensional one-hot vector indicating the reward location, with a single dummy action. Agent-B observes its grid position (also as a 16-dimensional one-hot vector) and selects from six actions: \( {\cal A}_B = \{\text{up}, \text{down}, \text{left}, \text{right}, \text{stand--still}, \text{dig}\} \).
Each agent can send one of 20 discrete messages per time step, enabling bidirectional communication. Episodes last up to 1,000 steps.

As this environment includes state transitions, we used IPPO-based methods for evaluation. Four conditions were compared: {\bf no-comm}, {\bf message}, {\bf cpc}, and {\bf shared}. In each setting, agents collected 1,024 time steps of experience per iteration using 8 parallel threads, with data split into four mini-batches for optimization. Training was run for \( 3 \times 10^6 \) time steps. Evaluation was performed periodically by averaging test performance. Metrics included group welfare \( W = \mathbb{E}[r_0 + r_1] \) and episode length \( \mathbb{E}[T] \), where expectations are empirical averages over test runs.

\subsubsection{Results}
\begin{figure}[t]
\centering
\includegraphics[width=\textwidth]{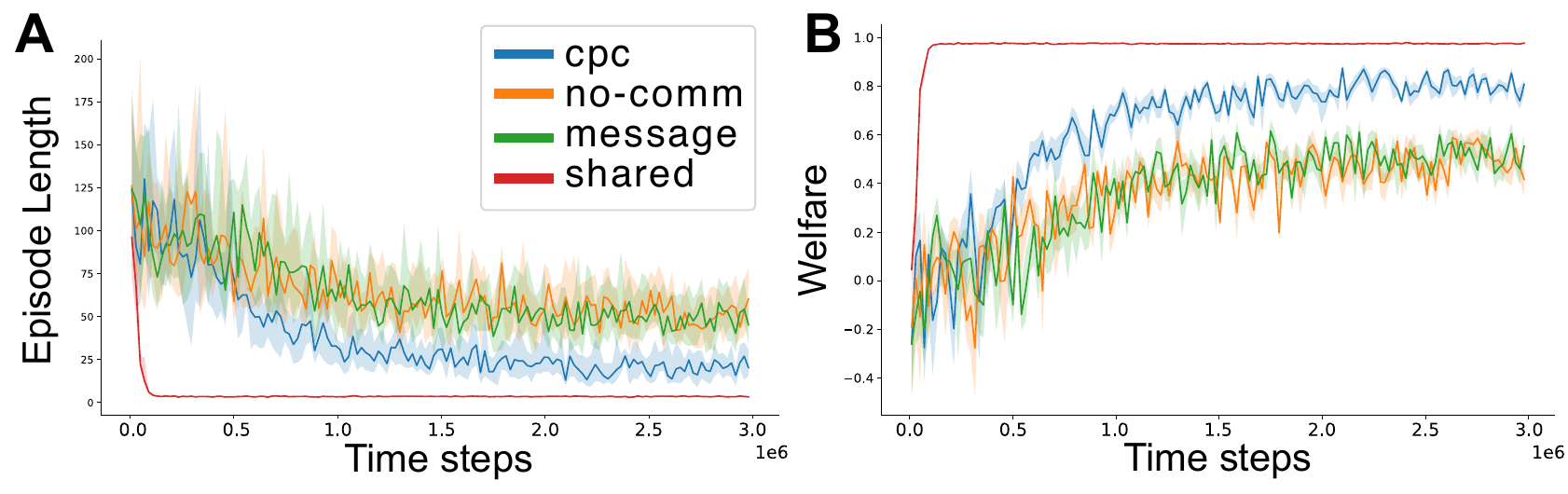}
\caption{Results in observer environment. {\bf A}) Episode length. {\bf B}) Group welfare.} 
\label{fig:r_obs}
\end{figure}

Figure~\ref{fig:r_obs} shows the experimental results. As in previous experiments, IQM and 95\% bootstrapped confidence intervals were computed over 10 trials. The {\bf shared} condition confirms that task performance improves with information sharing. Compared to the {\bf independent} and {\bf message} baselines, the {\bf cpc} condition shows significant gains across both metrics. An ablation study further evaluated the informativeness of messages generated by CPC agents for Agent-B through statistical testing.

\newpage
\section{Additional analysis}
\vspace{-2mm}
\subsection{Ablation Study}
\vspace{-2mm}

\begin{wrapfigure}[8]{r}{0.5\textwidth}
\vspace{-13mm}
\begin{center}
\includegraphics[width=0.5\textwidth]{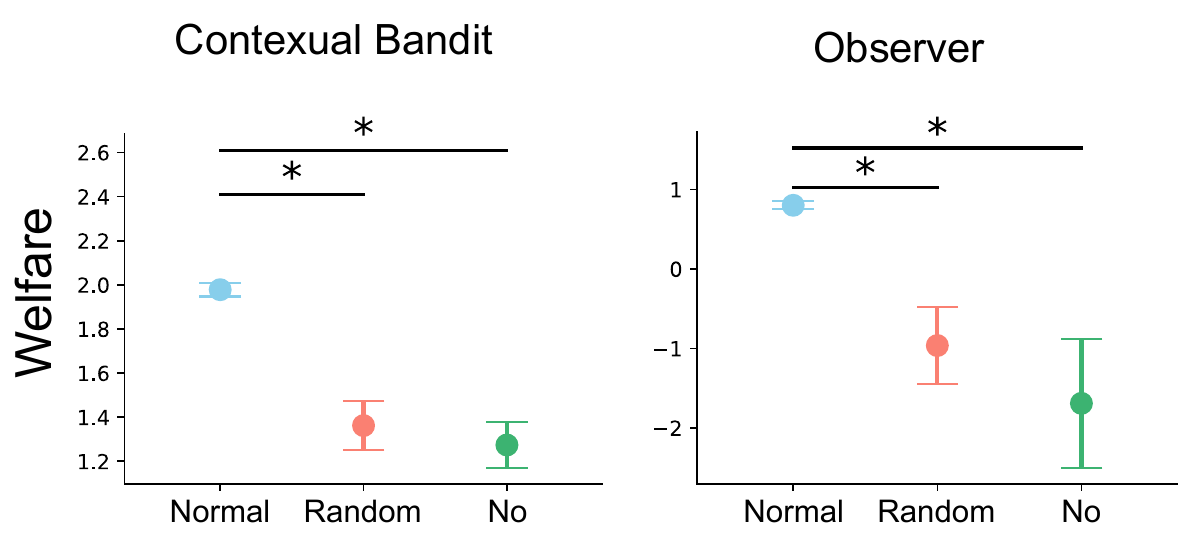}
\end{center}
\vspace{-7mm}
\caption{Overview of observer environment.} 
\label{fig:abl}
\end{wrapfigure}

The effectiveness of communication in MARL remains under debate~\cite{lowe2019pitfalls,peters2024survey}. To assess message utility in MARL-CPC, we conducted an ablation study to evaluate the impact of disrupting trained agents' messages. Two conditions were tested: {\bf random}, where messages were replaced with random values, and {\bf no}, where message vectors were set to zero when input as one-hot features.
We evaluated both interventions across 100 trials in the contextual bandit and observer environments. Results, summarized in Figure~\ref{fig:abl}, show a significant performance drop under both conditions, confirming that MARL-CPC establishes meaningful communication that contributes to task success.

\vspace{-2mm}
\subsection{Comparing in Cooperative Scenario}
\vspace{-2mm}
\begin{wrapfigure}[11]{r}{0.5\textwidth}
\centering
\vspace{-5mm}
\includegraphics[width=0.5\textwidth]{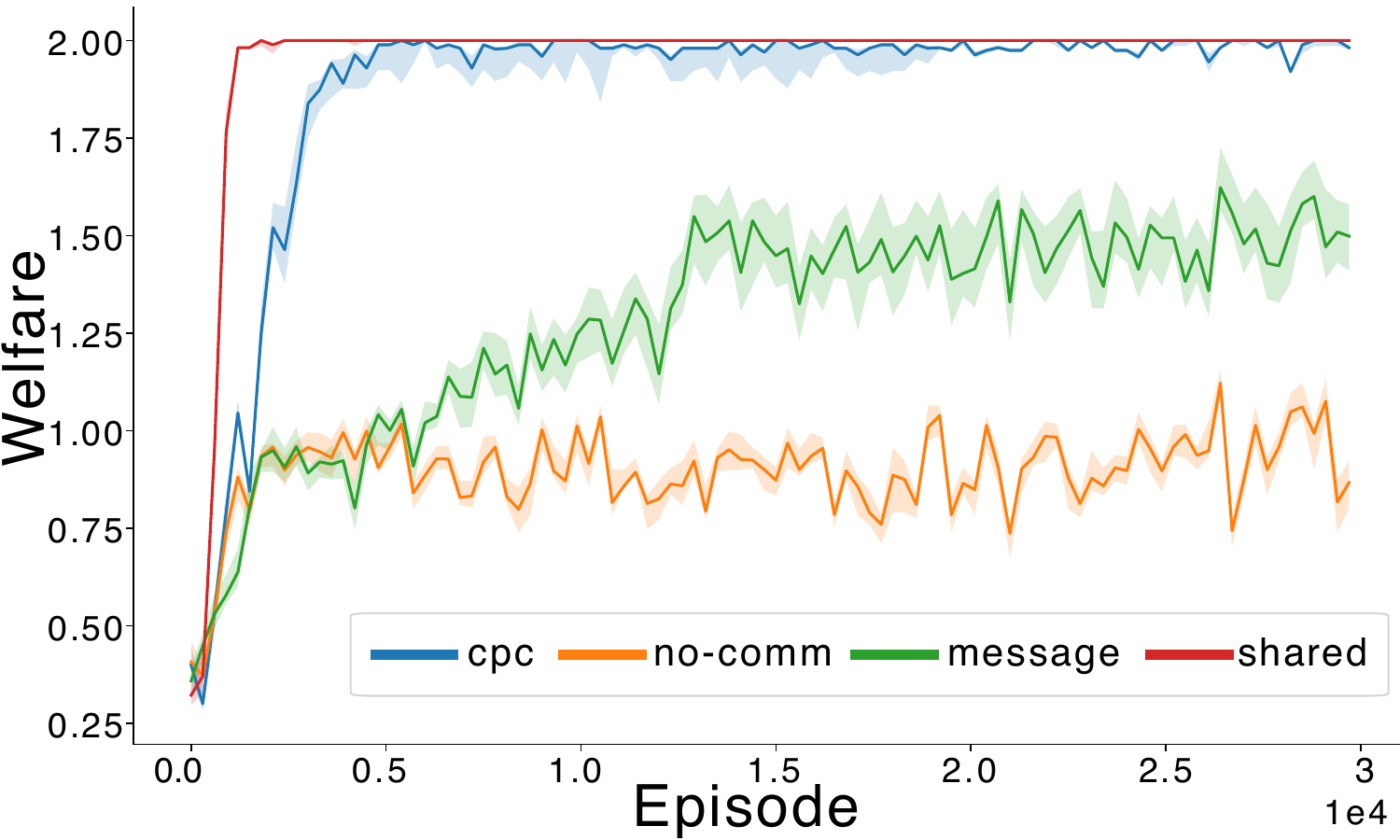}
\vspace{-8mm}
\caption{Results in cooperative environment.} 
\label{fig:r_coop}
\end{wrapfigure}
We examined communication learning under cooperative conditions by modifying the reward structure in the contextual bandit environment. Both agents now receive a reward of \( +1 \) only if they select the correct answer simultaneously; otherwise, they receive \( -0.1 \). Thus, an agent benefits from sharing information only when it knows the correct answer. This setting represents a cooperative MARL scenario.

Figure~\ref{fig:r_coop} presents IQM and 95\% bootstrapped confidence intervals over 10 runs for four conditions. As expected, the {\bf cpc} and {\bf shared} conditions achieve the highest rewards, while {\bf no-comm} fails to learn cooperative behavior. The {\bf message} condition, previously shown to enable communication in cooperative settings, also performs well, but learns more slowly and achieves lower final performance than {\bf cpc}. This likely reflects a fundamental difference: CPC integrates communication into representation learning as inference of global state, whereas the message-as-action paradigm treats communication as a learned action, requiring agents to jointly acquire both message generation and interpretation.

\vspace{-3mm}
\section{Conclusions}
\vspace{-2mm}
This study proposed MARL-CPC, a framework that applies CPC from emergent communication research to on-policy RL. MARL-CPC formulated the emergence of communication as a form of representation learning, based on a pseudo-joint generative modeling of multiple agents. This formulation was then decomposed into an objective function that each agent could optimize independently. The framework enabled agents to establish communication independently of the reward-driven mechanisms typical in conventional RL, and demonstrated improved performance in non-cooperative environments—settings that traditional MARL with communication has struggled to address effectively.


\vspace{-2mm}
\begin{credits}
\subsubsection{\ackname} 
This research was supported by the Japan Society for the Promotion of Science Grant in Aid for Transformative Research Areas (A) (23H04835). We thank Masatoshi Nagano, Nguyen Le Hoang, Noburo Saji, and Moe Ohkuma for the discussion. Figures are partially created with BioRender.com.

\vspace{-2mm}
\subsubsection{\discintname}
The authors have no competing interests to declare that are
relevant to the content of this article.
\end{credits}

%
%
%
\vspace{-2mm}
\bibliographystyle{splncs04}
\bibliography{reference}
\end{document}